\documentclass
[superscriptaddress,secnumarabic,amssymb,amsmath,nobibnotes,aps,prd,showkeys,showpacs,nofootinbib,onecolumn,12pt]{revtex4}%
\usepackage{graphicx}
\usepackage{epsf}
\usepackage{bm}
\usepackage{amsmath}
\usepackage{amsfonts}
\usepackage{amssymb}%
\providecommand{\U}[1]{\protect\rule{.1in}{.1in}}

\newcommand{\be}{\begin{equation}}
\newcommand{\ee}{\end{equation}}

\newcommand{\mincir}{\raise
-3.truept\hbox{\rlap{\hbox{$\sim$}}\raise4.truept\hbox{$<$}\ }}
\newcommand{\magcir}{\raise
-3.truept\hbox{\rlap{\hbox{$\sim$}}\raise4.truept\hbox{$>$}\ }}

\begin{document}
\title{Cosmological evolution in $f(T,B)$ gravity}
\author{Andronikos Paliathanasis}
\email{anpaliat@phys.uoa.gr}
\affiliation{Institute of Systems Science, Durban University of Technology, Durban 4000,
South Africa }
\affiliation{Instituto de Ciencias F\'{\i}sicas y Matem\'{a}ticas, Universidad Austral de
Chile, Valdivia, Chile}
\author{Genly Leon}
\email{genly.leon@ucn.cl}
\affiliation{Departamento de Matem\'{a}ticas, Universidad Cat\'{o}lica del Norte, Avda.
Angamos 0610, Casilla 1280 Antofagasta, Chile}

\begin{abstract}
For the fourth-order teleparallel $f\left(  T,B\right)  $ theory of gravity,
we investigate the cosmological evolution for the universe in the case of a
spatially flat Friedmann--Lema\^{\i}tre--Robertson--Walker background space.
We focus on the case for which $f\left(  T,B\right)  $ is separable, that is,
$f\left(  T,B\right)  _{,TB}=0$ and $f\left(  T,B\right)  $ is a nonlinear
function on the scalars $T$ and $B$. For this fourth-order theory we use a
Lagrange multiplier to introduce a scalar field function which attributes the
higher-order derivatives. In order to perform the analysis of the dynamics we
use dimensionless variables which allow the Hubble function to change sign.
The stationary points of the dynamical system are investigated both in the
finite and infinite regimes. The physical properties of the asymptotic
solutions and their stability characteristics are discussed.

\end{abstract}
\date{\today}
\maketitle

\section{Introduction}

\label{sec1}

Modified and extended theories of gravity have drawn the attention of
cosmologists in recent years because they provide a systematic and geometric
approach for the explanation of the cosmological observations
\cite{sp1,sp2,sp3,fg5,fg6,fg7,fg10,fg10a,fg10b}. In the family of modified
theories of gravity known as $f$-theories the gravitational Action Integral is
defined to be a function $f$ of specific geometric invariants
\cite{fg2,fg3,fg4,fg11,fg12,fg14,fg15,fg16,fg17,fg18,fg19,fg20}. The simplest
$f$-theory is the $f\left(  R\right)  $ gravity~\cite{Buda} in which the
Einstein-Hilbert Action, with or without the cosmological constant, is
recovered when $f\left(  R\right)  $ is a linear function. The main
characteristic of $f$-theories is that the new modified field equations
provide geometrodynamical components which drive the dynamics in order to
explain the cosmological evolution, while General Relativity should be recovered.

In this study we are interested in an extended higher-order teleparallel
$f$-theory known as $f\left(  T,B\right)  $ theory \cite{bahamonde}, where $T$
is the scalar function of the Weitzenb\"{o}ck connection and $B$ is the
boundary term which is the difference between $T$ and Ricci scalar
$R~$\cite{myr11}. The essential variables of teleparallel theory are the
vierbein fields, and scalar function $T$ is defined by first derivatives.
Consequently, a linear $f\left(  T,B\right)  =f_{1}T+f_{2}B$ theory is a
second-order theory. It describes the teleparallel equivalence of General
Relativity \cite{ein28,Hayashi79}. Furthermore, any $f\left(  T,B\right)
=F\left(  T\right)  +f_{2}B$ is also a second-order theory known as $F\left(
T\right)  $ theory \cite{Ferraro}. In the following we consider that $f\left(
T,B\right)  $ is separable, i.e.~ $f\left(  T,B\right)  _{,TB}=0$ and
$f\left(  T,B\right)  $ is a nonlinear function of $T$ and $B$, i.e.
$f_{,TT}\neq0$ and $f_{,BB}\neq0$. Hence, the resulting theory is a
fourth-order theory because $B$ includes second derivatives of the vierbein fields.

In \cite{ftb01} the authors investigated the gravitomagnetism characteristic
phenomena in the weak field in $f\left(  T,B\right)  $ theory. Moreover,
cosmological bouncing solutions were studied in \cite{ftb02} while the
cosmological perturbations of the theory were investigated in \cite{ftb02a}.
The linear function $F$ in the scalar function $T~$\ was the subject of study in
\cite{anprd} where a scalar field description was used in order to write the field
equations as second-order differential equations. The theory admits a
minisuperspace description which was applied to study the quantization and the
Wheeler-DeWitt equation \cite{an11}. Some other studies of $f\left(  T,B\right)
$ theory which we discuss in the following sections are
\cite{anjcap,angrg,cel1,cel2}. The plan of the paper is as follows.

In Section \ref{sec2}, we present the cosmological model of our consideration.
We consider $f\left(  T,B\right)  $ teleparallel theory where $f\left(
T,B\right)  =F\left(  T\right)  +\beta\left(  B\right)  $ and $F\left(
T\right)  ,~\beta\left(  B\right)  $ are nonlinear functions. \ In addition,
for the background space, we assume a spatially flat
Friedmann--Lema\^{\i}tre--Robertson--Walker (FLRW) spacetime. For this
cosmological model the field equations are of fourth-order in the scale
factor. However, we define a new scalar field which attributes the additional
degrees of freedom such that to write the field equations as second-order
equations by increasing the number of dependent variables. Hence, the
dependent variables are the scalar factor and the new scalar field. In Section
\ref{sec3} we present the main results of our analysis. We perform a detailed
analysis of the dynamics for the cosmological field equations by using
dimensionless variables. Because of the geometric terms in the field equations
which correspond to the higher-order derivatives, we observe that the Hubble
function can change sign during the cosmological evolution. Hence, a new set
of variables, different from those of the Hubble normalization, are defined which
allow the Hubble function to be continuous in whole range of values. We
investigate the stationary points and their stability properties for the field
equations in the new variables. Every stationary point corresponds to a
specific epoch in the cosmological history and provides us with important
information about the cosmological evolution and the viability of the theory.
The stationary points are studied in the finite and the infinite regimes.
Finally, in Section \ref{sec4} we summarize our results and we draw our conclusions.

\section{$f\left(  T,B\right)  $ cosmology}

\label{sec2}

In this work we consider the fourth-order generalized teleparallel theory of
gravity known as $f\left(  T,B\right)  $ theory. Contrary to General
Relativity, in teleparallel theories the dynamical variables are the vierbein
fields. The later form an orthonormal basis for the tangent space at each
point, $x^{\mu}$, of the manifold,$~g(e_{i},e_{j})=\mathbf{e}_{i}%
\cdot\mathbf{e}_{i}=\eta_{ij}$, where $\eta_{ij}~$is the four-dimensional
Minkowski spacetime. The equivalent form in the coordinate system%

\begin{equation}
g_{\mu\nu}(x)=\eta_{ij}h_{\mu}^{i}(x)h_{\nu}^{j}(x)~,~e_{i}=h_{i}^{\mu}\left(
x\right)  \partial_{i}.
\end{equation}

By definition, the fundamental connection in teleparallel theories is the
Weitzenb\"{o}ck connection $\hat{\Gamma}^{\lambda}{}_{\mu\nu}=h_{a}^{\lambda
}\partial_{\mu}h_{\nu}^{a}$, from where we can define the nonnull torsion
tensor, \cite{ftt0,ftt1}
\begin{equation}
T_{\mu\nu}^{\beta}=\hat{\Gamma}_{\nu\mu}^{\beta}-\hat{\Gamma}_{\mu\nu}^{\beta
}=h_{i}^{\beta}(\partial_{\mu}h_{\nu}^{i}-\partial_{\ qu}h_{\mu}^{i}),
\end{equation}
with scalar
\[
T={S_{\beta}}^{\mu\nu}{T^{\beta}}_{\mu\nu},~
\]
in which ${S_{\beta}}^{\mu\nu}~$is defined as ${S_{\beta}}^{\mu\nu}=\frac
{1}{2}({K^{\mu\nu}}_{\beta}+\delta_{\beta}^{\mu}{T^{\theta\nu}}_{\theta
}-\delta_{\beta}^{\nu}{T^{\theta\mu}}_{\theta}).~T$ is the scalar function
which is used for the definition of the gravitational Action
Integral.\ The quantity ${K^{\mu\nu}}_{\beta}$ is the contorsion tensor and equals
the difference between the Levi-Civita connections in the holonomic and the
nonholonomic frame.~${K^{\mu\nu}}_{\beta}~$is defined as ${K^{\mu\nu}}_{\beta
}=-\frac{1}{2}({T^{\mu\nu}}_{\beta}-{T^{\nu\mu}}_{\beta}-{T_{\beta}}^{\mu\nu
}).$

In $f\left(  T,B\right)  $ gravity the gravitational Action Integral is a
function $f$ of the scalar $T$ and of the boundary term $B$ which is the boundary
term $B=2e_{\nu}^{-1}\partial_{\nu}\left(  eT_{\rho}^{~\rho\nu}\right)  $,
that is, \cite{bahamonde}%

\begin{equation}
S\equiv\frac{1}{16\pi G}\int d^{4}xef\left(  T,B\right)  , \label{ftb.01}%
\end{equation}
in which $e=\det(e_{\mu}^{i})=\sqrt{-g}.~$Because $B=R+T$, where$~R$ is the
Ricci scalar, the Action Integral\ (\ref{ftb.01}) is equivalent to \cite{anprd}%
\[
S\equiv\frac{1}{16\pi G}\int d^{4}xe\bar{f}\left(  T,R\right)  ,~\text{or~}%
S\equiv\frac{1}{16\pi G}\int d^{4}xe\hat{f}\left(  R,B\right)  ~,
\]
in which $f\left(  T,B\right)  =\bar{f}\left(  T,R\right)  $ and $f\left(
T,B\right)  =\hat{f}\left(  R,B\right)  $.

The field equations follow from the variation of (\ref{ftb.01}) with respect
to the vierbein fields,~\cite{bahamonde}
\begin{align}
0  &  =\frac{1}{2}eh_{a}^{\lambda}\left(  f_{,B}\right)  ^{;\mu\nu}g_{\mu\nu
}-\frac{1}{2}eh_{a}^{\sigma}\left(  f_{,B}\right)  _{;\sigma}^{~~~;\lambda
}+\frac{1}{4}e\left(  Bf_{,B}-\frac{1}{4}f\right)  h_{a}^{\lambda}\,+(eS_{a}%
{}^{\mu\lambda})_{,\mu}f_{,T}\nonumber\\
&  ~\ ~+e\left(  (f_{,B})_{,\mu}+(f_{,T})_{,\mu}\right)  S_{a}{}^{\mu\lambda
}~-ef_{,T}T^{\sigma}{}_{\mu a}S_{\sigma}{}^{\lambda\mu}, \label{ftb.02}%
\end{align}
where \textquotedblleft$;$\textquotedblright\ $\ $ denotes the covariant
derivative. The field equations (\ref{ftb.02}) for a linear function $f\left(
T,B\right)  =f_{1}T+f_{2}B$ reduce to that of Einstein teleparallel theory of
gravity, which is equivalent with General Relativity, while, if it is linear, to
$B$, i.e. $f\left(  T,B\right)  =F\left(  T\right)  +f_{2}B$, the second-order
$F\left(  T\right)  $ teleparallel gravity is recovered. In addition, when
$f\left(  T,B\right)  =f\left(  T-B\right)  ,$ i.e. $f\left(  T,B\right)
=f\left(  R\right)  $, the field equations (\ref{ftb.02}) are those of the
fourth-order theory known as $f\left(  R\right)  $ theory.

With the use of the Einstein tensor, $G_{a}^{\lambda}$, the field equations
(\ref{ftb.02}) can be written in the equivalent form%
\begin{equation}
eG_{a}^{\lambda}=G_{eff}\left(  \mathcal{T}_{a}^{\left(  T\right)  }%
{}^{\lambda}+\mathcal{T}_{a}^{\left(  B\right)  }{}{}^{\lambda}\right)  ,
\label{ftb.03}%
\end{equation}
in which $G_{eff}=\frac{4\pi G}{f_{,T}}$ is the effective varying
gravitational constant and~$\mathcal{T}_{a}^{\left(  T\right)  }{}^{\lambda},$ and
$\mathcal{T}_{a}^{\left(  B\right)  }{}{}^{\lambda}$ are the effective energy
momentum tensors \cite{anprd}
\begin{equation}
4\pi Ge\mathcal{T}_{a}^{\left(  T\right)  }{}^{\lambda}=-\left[  \frac{1}%
{4}\left(  Tf_{,T}-f\right)  eh_{a}^{\lambda}+e(f_{,T})_{,\mu}S_{a}{}%
^{\mu\lambda}\right]  , \label{ftb.04}%
\end{equation}%
\begin{equation}
4\pi Ge\mathcal{T}_{a}^{\left(  B\right)  }{}^{\lambda}=-\left[
e(f_{,B})_{,\mu}S_{a}{}^{\mu\lambda}-\frac{1}{2}e\left(  h_{a}^{\sigma}\left(
f_{,B}\right)  _{;\sigma}^{~~~;\lambda}-h_{a}^{\lambda}\left(  f_{,B}\right)
^{;\mu\nu}g_{\mu\nu}\right)  +\frac{1}{4}eBf_{,B}h_{a}^{\lambda}\right]  .
\label{ftb.05}%
\end{equation}

In this work we are interested in the case for which $f\left(  T,B\right)  =F\left(
T\right)  +\beta\left(  B\right)  $. In such consideration, the effective
energy momentum tensors are written as \cite{anprd}%
\begin{equation}
4\pi Ge\mathcal{T}_{a}^{\left(  T\right)  }{}^{\lambda}=-\left[  \frac{1}%
{4}\left(  TF_{,T}-F\right)  eh_{a}^{\lambda}+e(F_{,TT})T_{,\mu}S_{a}{}%
^{\mu\lambda}\right]  \label{ftb.12}%
\end{equation}%
\begin{align}
4\pi Ge\mathcal{T}_{a}^{\left(  B\right)  }{}^{\lambda}  &  =-e(\beta\left(
B\right)  _{,BB})B_{;\mu}S_{a}{}^{\mu\lambda}-\frac{1}{4}e\left(
B\beta\left(  B\right)  _{,B}-\beta\left(  B\right)  \right)  h_{a}^{\lambda
}\label{ftb.13}\\
&  +\frac{1}{2}e\left(  h_{a}^{\sigma}\left(  \beta\left(  B\right)
_{,B}\right)  _{;\sigma}^{~~~;\lambda}-h_{a}^{\lambda}\left(  \beta\left(
B\right)  _{,B}\right)  ^{;\mu\nu}g_{\mu\nu}\right) \nonumber
\end{align}
while the varying gravitational constant is $\frac{G_{eff}}{4\pi G}=\left(
F_{,T}\right)  ^{-1}.~$\ In such a case, in $\frac{\partial\mathcal{T}%
_{a}^{\left(  T\right)  }{}^{\lambda}}{\partial B}=0$ and ${}\frac
{\partial\mathcal{T}_{a}^{\left(  B\right)  }{}^{\lambda}}{\partial T}=0,$
which we can say that there is no interaction in the Action Integral
between the two effective fluids.

Moreover, in such a case the energy-momentum tensor which includes the
higher-order derivatives of the $f\left(  T,B\right)  $ theory can be written
in an equivalent form by using a scalar field. Indeed, if we define
$\phi=\beta\left(  B\right)  _{,B}$, then (\ref{ftb.13}) is simplified as
\cite{anjcap}
\begin{equation}
4\pi Ge\mathcal{T}_{a}^{\left(  B\right)  }{}^{\lambda}=-\left[  e\phi_{;\mu
}S_{a}{}^{\mu\lambda}-\frac{1}{2}e\left(  h_{a}^{\sigma}\left(  \phi\right)
_{;\sigma}^{~~~;\lambda}-h_{a}^{\lambda}\left(  \phi\right)  ^{;\mu\nu}%
g_{\mu\nu}\right)  -\frac{1}{4}eV\left(  \phi\right)  h_{a}^{\lambda}\right]
, \label{ftb.14}%
\end{equation}
where $V\left(  \phi\right)  =\left(  \beta\left(  B\right)  -B\beta\left(
B\right)  _{,B}\right)  $.

This is not a canonical scalar field and the theory does not belong to the
scalar tensor theories. However, there is a direct relation with the
teleparallel dark energy models \cite{cd1} under the action of conformal
transformations \cite{ww1}.

\subsection{FLRW background space}

According to the cosmological principle for the background space we consider a
spatially flat FLRW spacetime described by the scale factor $a\left(
t\right)  $ with line element%
\begin{equation}
ds^{2}=N^{2}\left(  t\right)  dt^{2}-a^{2}\left(  t\right)  \left(
dx^{2}+dy^{2}+dz^{2}\right)  , \label{ftb.15}%
\end{equation}
where $N\left(  t\right)  $ is the lapse function. For the vierbein we
consider the following diagonal frame%
\begin{equation}
h_{\mu}^{i}(t)=\text{diag}\left(  N\left(  t\right)  ,a\left(  t\right)  ,a\left(
t\right)  ,a\left(  t\right)  \right)  . \label{ftb.150}%
\end{equation}

In this case the field equations (\ref{ftb.02}) are described by the
point-like Lagrangian \cite{anprd}
\begin{equation}
L\left(  N,a,\dot{a},T,B,\dot{B}\right)  =-\frac{6}{N}aF_{,T}\dot{a}^{2}%
+\frac{6}{N}a^{2}\dot{a}\dot{\beta}\left(  B\right)  _{,B}+Na^{3}\left(
F-TF_{,T}\right)  +Na^{3}\left(  \beta\left(  B\right)  -B\beta\left(
B\right)  _{,B}\right)  \label{ft.01}%
\end{equation}
or equivalently with the use of the scalar field $\phi,~$%
\begin{equation}
L\left(  N,a,\dot{a},T,B,\dot{B}\right)  =-\frac{6}{N}aF_{,T}~\dot{a}%
^{2}+\frac{6}{N}a^{2}\dot{a}\dot{\phi}+Na^{3}\left(  F-TF_{,T}\right)
+Na^{3}V\left(  \phi\right)  \label{ft.02}%
\end{equation}

The field equations for $N=1$ are written as%
\begin{equation}
6F_{,T}H^{2}-6H\dot{\phi}+\left(  F-TF_{,T}\right)  +V\left(  \phi\right)
=0~, \label{ft.03}%
\end{equation}%
\begin{equation}
2F_{,T}\left(  2\dot{H}+3H^{2}\right)  +4F_{TT}\dot{T}H+\left(  F-TF_{,T}%
\right)  +V\left(  \phi\right)  -2\ddot{\phi}=0~, \label{ft.04}%
\end{equation}%
\begin{equation}
6\left(  \dot{H}+3H^{2}\right)  +V_{,\phi}\left(  \phi\right)  =0~,
\label{ft.05}%
\end{equation}%
and
\begin{equation}
T+6H^{2}=0,~ \label{ft.06}%
\end{equation}
where $H=\frac{\dot{a}}{a}$ is the Hubble function.

Equation (\ref{ft.03}) is the first modified Friedmann's equation, known as
the constraint equation. Expression (\ref{ft.04}) is the second modified
Friedmann's equation, while expressions (\ref{ft.05}) and (\ref{ft.06}) are
the constraint equations for the scalars$~T$ and $B$.

In the following, we consider $F\left(  T\right)  =T+\Lambda\left(  T\right)
$ in which $\Lambda\left(  T\right)  =\Lambda_{0}T^{n}~$\cite{ftn1,ftn2}.
Therefore, the field equations (\ref{ft.03})-(\ref{ft.04}) are written as%
\begin{equation}
6H^{2}-6H\dot{\phi}+\left(  \Lambda-2T\Lambda_{,T}\right)  +V\left(
\phi\right)  =0~, \label{ft.07}%
\end{equation}%
and
\begin{equation}
2\left(  2\dot{H}+3H^{2}\right)  +4\left(  \Lambda_{,T}-2\Lambda
_{,TT}T\right)  \dot{H}+\left(  \Lambda-2T\Lambda_{,T}\right)  +V\left(
\phi\right)  -2\ddot{\phi}=0~, \label{ft.08}%
\end{equation}
or
\begin{equation}
H^{2}-H\dot{\phi}+\frac{\Lambda_{0}}{6}\left(  1-2n\right)  T^{n}+\frac{1}%
{6}V\left(  \phi\right)  =0~, \label{ft.09}%
\end{equation}%
and
\begin{equation}
2\left(  2\dot{H}+3H^{2}\right)  +4n\Lambda_{0}\left(  3-2n\right)
T^{n-1}\dot{H}+\Lambda_{0}\left(  1-2n\right)  T^{n}+V\left(  \phi\right)
-2\ddot{\phi}=0~. \label{ft.10}%
\end{equation}

As far as the scalar field potential $V\left(  \phi\right)  $ is concerned, we
assume the exponential potential $V\left(  \phi\right)  =V_{0}e^{-\lambda\phi
}$ which correspond, to the $\beta\left(  B\right)  =-\frac{1}{\lambda}%
B\ln\left(  B\right)  $ theory \cite{anjcap}. That function has been widely
studied before in the case of $f\left(  T,B\right)  =T+\beta\left(  B\right)
$ cosmology. It provides scaling solutions for $\lambda\neq3$ and de Sitter
solution for $\lambda=3~$\cite{anjcap}. Of course other functional forms for
the scalar field potential can be considered, as power-law functions which
correspond to power-law $\beta\left(  B\right)  $, but, as it has been shown in
\cite{angrg}, the exponential potential provides many important results which
are related to inflation and which do not exist for nonexponential potential,
similarly with the analysis for the quintessence field \cite{copeland,anqq}.

\section{Dynamical analysis}

\label{sec3}

In the following we perform a detailed study of the dynamics for the
gravitational field equations under consideration. Such an analysis is important
in order to understand the general evolution of the cosmological evolution and
to infer about the viability of the theory. This method has been applied in
various cosmological theories with many interesting results, see for instance
\cite{dn1,dn2,dn3,dn4,dn5,dn6,dn7,dn8} and references therein. For $n=1$, or
$\Lambda_{0}=0$, the complete analysis of the dynamics was performed in
\cite{angrg}. Recently, in \cite{cel1,cel2} separable$~f\left(  T,B\right)
=F\left(  T\right)  +\beta\left(  B\right)  $ models with nonlinear functions
$F\left(  T\right)  $ or $\beta\left(  B\right)  $ were investigated.  However,
the authors considered a power-law function for $\beta\left(  B\right)  $ which,
as we discussed above, does not provide a cosmological history as the model
$\beta\left(  B\right)  =-\frac{1}{\lambda}B\ln\left(  B\right)  $. Moreover
in \cite{cel1} the authors selected to work on the so-called $H$-normalization
\cite{copeland}.\ While the analysis in \cite{cel1} is correct, it is not
complete because of the selection of the variables. From (\ref{ft.07}) it is
clear that the Hubble function can change sign and be zero, hence a new
parametrization which allows the change of the sign for the Hubble
parameter is necessary to be considered.

We define the new dimensionless variables \cite{dn8},%
\begin{equation}
x=\frac{\dot{\phi}}{\sqrt{1+H^{2}}}~,~z=\frac{T^{n}}{1+H^{2}}~,~y=\frac
{V\left(  \phi\right)  }{1+H^{2}}~,\eta=\frac{H}{\sqrt{1+H^{2}}},
\label{ft.11}%
\end{equation}
which differ from that of the $H$-normalization. Variable $\eta$ indicates the
sign of $H$ and, when $H=0$, $\eta=0$ follows . Moreover, when $H$ takes
values near to infinity, $\left\vert \eta\right\vert =1$.

In the new dimensionless variables the gravitational field equations are%
\begin{align}
18\frac{dx}{d\omega}  &  =3\text{$\Lambda$}_{0}(2n-1)z\left(  \eta
^{3}(3-2\lambda)+\eta(-2(\lambda-3)n-3)+2\lambda nx+\eta^{4}\lambda x\right)
+\nonumber\\
&  +\eta\lambda\text{$\Lambda$}_{0}^{2}(1-2n)^{2}nz^{2}+18\eta^{2}\left(
-6\eta+3\left(  \eta^{2}+1\right)  x+\lambda(x-\eta)(\eta x-2)\right)  ,
\label{ft.12}%
\end{align}%
\begin{equation}
\frac{dz}{d\omega}=\frac{1}{3}\eta z\left(  \eta^{2}-n\right)  (-6\eta
(\lambda-3)+\eta\lambda\text{$\Lambda$}_{0}(2n-1)z+6\lambda x), \label{ft.13}%
\end{equation}%
and
\begin{equation}
\frac{d\eta}{d\omega}=\frac{1}{6}\eta^{2}\left(  \eta^{2}-1\right)
(-6\eta(\lambda-3)+\eta\lambda\text{$\Lambda$}_{0}(2n-1)z+6\lambda x),
\label{ft.14}%
\end{equation}
where from the constraint equation (\ref{ft.09}) there follows the algebraic
equation
\begin{equation}
y=\eta(\eta(\text{$\Lambda$}_{0}(2n-1)z-6)+6x). \label{ft.15}%
\end{equation}
The new independent parameter, $\omega$, is defined as~$dt=\eta\sqrt{1-\eta
^{2}}d\omega$. Moreover, parameter $\lambda=-\frac{V_{,\phi}}{V}$, where for
the exponential potential of our consideration $V\left(  \phi\right)
=V_{0}e^{-\lambda_{0}\phi}$, $\lambda=\lambda_{0}$ is always a constant.

However,the  parameter $z$ is not independent. Indeed, $z=\left(  -6\right)
^{n}\eta^{2n}\left(  1-\eta^{2}\right)  ^{1-n}$. Consequently, we end with a
two-dimensional system, equations (\ref{ft.12}), (\ref{ft.14})

Every stationary point of the dynamical system (\ref{ft.12}), (\ref{ft.14})
describes a specific exact solution for the field equations. At the stationary
points $P$, the effective cosmological fluid has an effective equation of
state parameter $w_{eff}=w_{eff}\left(  P\right)  $, where $w_{eff}%
=-1-\frac{2}{3}\frac{\dot{H}}{H^{2}}$.\ Hence the scale factor of the FLRW
spacetime is calculated to be $a\left(  t\right)  =a_{0}t^{\frac{2}{3\left(
1+w_{eff}\right)  }}$, for $w_{eff}\neq-1$ or $a\left(  t\right)
=a_{0}e^{H_{0}t}$,~for $w_{eff}=-1$. In the new dimensionless variables the
$w_{eff}$ is expressed as%
\[
w_{eff}=1+\frac{\lambda}{9}\frac{y}{\eta^{2}}
\]
or, equivalently,%
\begin{equation}
w_{eff}\left(  x,z,\eta\right)  =\frac{1}{9}\left(  9-6\lambda-\left(
1-2n\right)  \lambda\Lambda_{0}z+6\lambda\frac{x}{\eta}\right)  .
\label{ft.16}%
\end{equation}

Moreover, it is important to study the stability of the stationary points.
Such an analysis provides important information for the evolution of the
universe near to the asymptotic solutions as also, to find the final evolution
of the system.

\subsection{Stationary points}

The stationary points $P=\left(  x\left(  P\right)  ,\eta\left(  P\right)
\right)  $ of the dynamical system (\ref{ft.12}), (\ref{ft.13}), (\ref{ft.14})
are derived to be%
\begin{align*}
P_{1}  &  =\left(  1,1\right)  ~,~P_{2}=(-1,-1)~,~\\
P_{3}  &  =\left(  2-\frac{6}{\lambda},1\right)  ~,~P_{4}=\left(  \frac
{6}{\lambda}-2,-1\right)  ~,\\
P_{5}  &  =\left(  \frac{\left(  \lambda-3\right)  }{\lambda}\eta\left(
\eta^{2}-1\right)  ,~\eta\right)  ~,~z\left(  P_{5}\right)  =\frac{6\left(
\lambda-3\right)  }{\left(  2n-1\right)  }\eta^{2}.\\
P_{6}  &  =\left(  x,0\right)~.
\end{align*}

Points $P_{1},~P_{2}$ describe asymptotic solutions where only the kinetic
part of the scalar field $\phi$ contributes in the cosmological fluid. The
effective equation of state parameter has the value $w_{eff}\left(  P_{\left(
3,4\right)  }\right)  =1$, which means that the effective fluid is that of a
stiff fluid source and that the scale factor is approximated as $a\left(
t\right)  =a_{0}t^{\frac{1}{3}}$. From (\ref{ft.11}) and (\ref{ft.06}) points
are accepted for $n<0$.

Points $P_{3},~P_{4}$ describe solutions where the kinetic and potential parts
of the scalar field $\phi$, that is, the boundary term $B$, contributes in the
total solution. We calculate $w_{eff}\left(  P_{\left(  3,4\right)  }\right)
=\frac{2\lambda}{3}-3$, which means that the scale factor is $a\left(
t\right)  =a_{0}t^{\frac{1}{\lambda-3}}$, for $\lambda\neq-3$ and $a\left(
t\right)  =a_{0}e^{H_{0}t}$ for $\lambda=3$. In the latter case for
$\lambda=3$, it follows that $x\left(  P_{\left(  3,4\right)  }\right)  =0$,
that is, from (\ref{ft.15}) $y\left(  P_{\left(  3,4\right)  }\right)  =1.$
For $\lambda<4$, the stationary points describe accelerated universes,
$w_{eff}\left(  P_{\left(  3,4\right)  }\right)  <-\frac{1}{3}$, while for
$\lambda=\frac{9}{2}$ and $\lambda=5$, the dust fluid solution, $w_{eff}%
\left(  P_{\left(  3,4\right)  }\right)  =0,$ and the radiation solution,
$w_{eff}\left(  P_{\left(  3,4\right)  }\right)  =\frac{1}{3},$ are recovered
respectively. Similarly with before points are accepted for $n<0$.

Point $P_{5}$ describes a family of de Sitter universes, $w_{eff}\left(
P_{5}\right)  =-1$, where the scale factor is $a\left(  t\right)
=a_{0}e^{H_{0}\left(  P_{5}\right)  t}$. Hence, from (\ref{ft.11}) and
(\ref{ft.06}) we find~$T^{n}=\frac{6\left(  \lambda-3\right)  }{\left(
2n-1\right)  \lambda\Lambda_{0}}H\left(  P_{5}\right)  ^{2}$, that is,
$H\left(  P_{5}\right)  ^{2\left(  n-1\right)  }=-\frac{\left(  -6\right)
^{-\left(  n-1\right)  }\left(  \lambda-3\right)  }{\left(  2n-1\right)
\lambda\Lambda_{0}}$. Therefore for $\lambda=3$, we find that the point
reduces to $P_{6}$ for $x\left(  P_{6}\right)  =0$.

Finally, point $P_{6}$ describes Minkowski spacetimes \thinspace$H\left(
P_{\left(  6,7\right)  }\right)  =0$. However, at this point we should discuss
about the existence for this points. From (\ref{ft.11}) it follows
$T^{n}=~z\left(  1+H^{2}\right)  $. Thus, for Minkowski spacetime and from
(\ref{ft.06}) it follows that point $P_{6}$ is accepted for $n>0$.

\subsection{Stability of the stationary points}

We continue our analysis by investigate the stability properties for the
stationary points of the dynamical system. For simplicity of our calculations
and only for the presentation of the results we prefer to work with the
three-dimensional dynamical system (\ref{ft.12}), (\ref{ft.13}),
(\ref{ft.14}). In order to infer about the stability of the points we
determine the eigenvalues for the linearized system around the stationary points.

For the stationary points $P_{1},~P_{2}$ the eigenvalues are derived to be%
\[
e_{1}\left(  P_{\left(  1,2\right)  }\right)  =6~,~e_{2}\left(  P_{\left(
1,2\right)  }\right)  =6\left(  1-n\right)  \text{ and }e_{3}\left(
P_{\left(  1,2\right)  }\right)  =6-\lambda.
\]
Therefore, the points $P_{1},~P_{2}$ describe unstable asymptotic solutions.
For $n<1$ and $\lambda<6$ points are sources otherwise they are saddle points.

Furthermore, for the points $P_{3},~P_{4}$ we calculate%
\[
e_{1}\left(  P_{\left(  3,4\right)  }\right)  =\lambda-6~,~e_{2}\left(
P_{\left(  3,4\right)  }\right)  =2\left(  \lambda-3\right)  \text{ and }%
e_{3}\left(  P_{\left(  3,4\right)  }\right)  =-2\left(  \lambda-3\right)
\left(  n-1\right)  .
\]
We conclude that for $\lambda<3$ and $n<1$ the stationary points are
attractors and describe stable scaling solutions. For $n<1$ and $\lambda>6$
points are sources otherwise they are saddle points.

\begin{figure}[ptb]
\centering\includegraphics[width=1\textwidth]{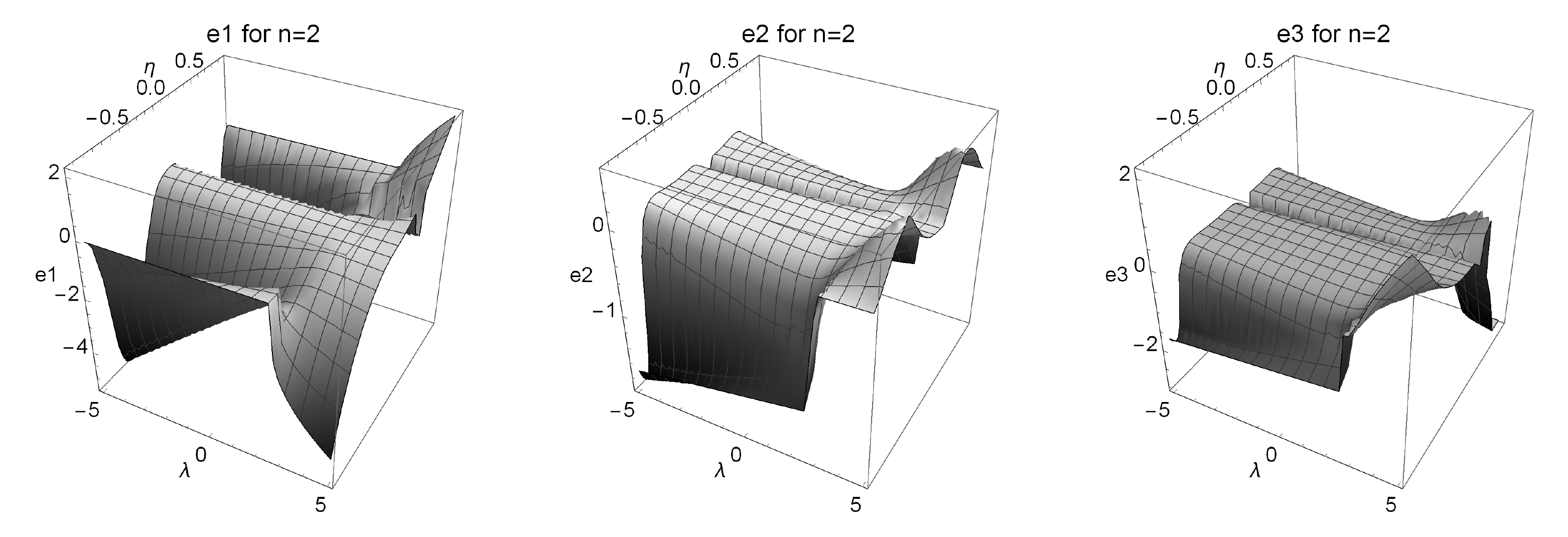} \caption{Qualitative
evolution for the real part for eigenvalues of the linearized system near to
point $P_{5}$. Plots correspond to the eigenvalues for $n=2$. We observe that
points $P_{5}~$are attractor for $\lambda<3$.}%
\label{ftb1}%
\end{figure}

As far as the eigenvalues of point $P_{5}$ are concerned, they depend on the
three parameters $\left\{  \lambda,\eta,n\right\}  $, where $\Lambda_{0}$ has
been replaced by the constraint found before. Because of the nonlinearity of
the eigenvalues we plot them numerically. In Fig. \ref{ftb1} we present the
qualitative evolution for the three eigenvalues $e_{1}\left(  P_{5}\right)
~,~e_{2}\left(  P_{5}\right)  $\ and $e_{3}\left(  P_{5}\right)  .$ The plots
are in the space of variables $\left\{  \lambda,\eta\right\}  $ for $n=2$ and
for $\lambda\neq3$. From Fig. \ref{ftb1} it is obvious that points $P_{5}$ can
be attractors.

For point $P_{6}$ the eigenvalues of the linearized system are all equal to
zero, that is,$~e_{1}\left(  P_{6}\right)  =0,~e_{2}\left(  P_{6}\right)
=0$\ and $e_{3}\left(  P_{6}\right)  =0$. In order to infer on the stability
of the stationary points the CMT should be applied. However, in this work we
prefer to present the phase-space diagrams where we find that the points are saddle.

\begin{figure}[ptb]
\centering\includegraphics[width=1\textwidth]{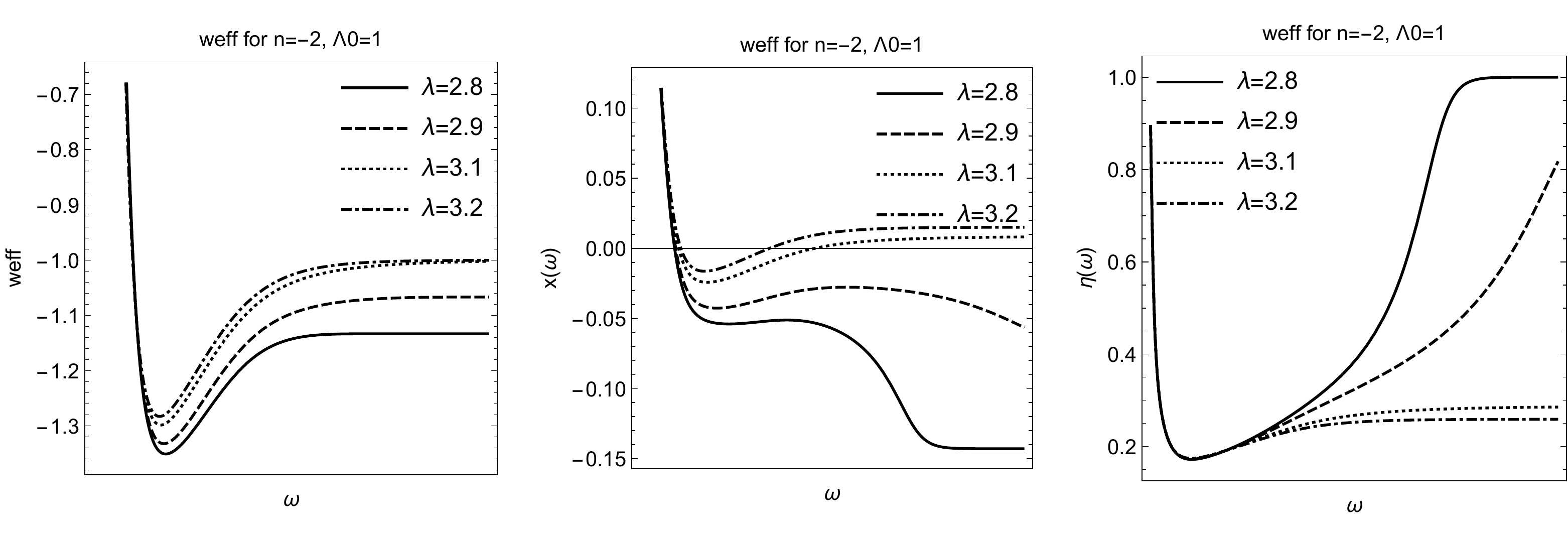} \caption{Qualitative
evolution for the physical variable $w_{eff}$ and for the parameters $x,~\eta$
for $n=-2,~\Lambda_{0}=1$ and for different values of $\lambda$. We observe
that the final attractor is point $P_{3}$ for $\lambda<3$, and $P_{5}$ for
$\lambda>3$.}%
\label{ftb2}%
\end{figure}

\begin{figure}[ptb]
\centering\includegraphics[width=1\textwidth]{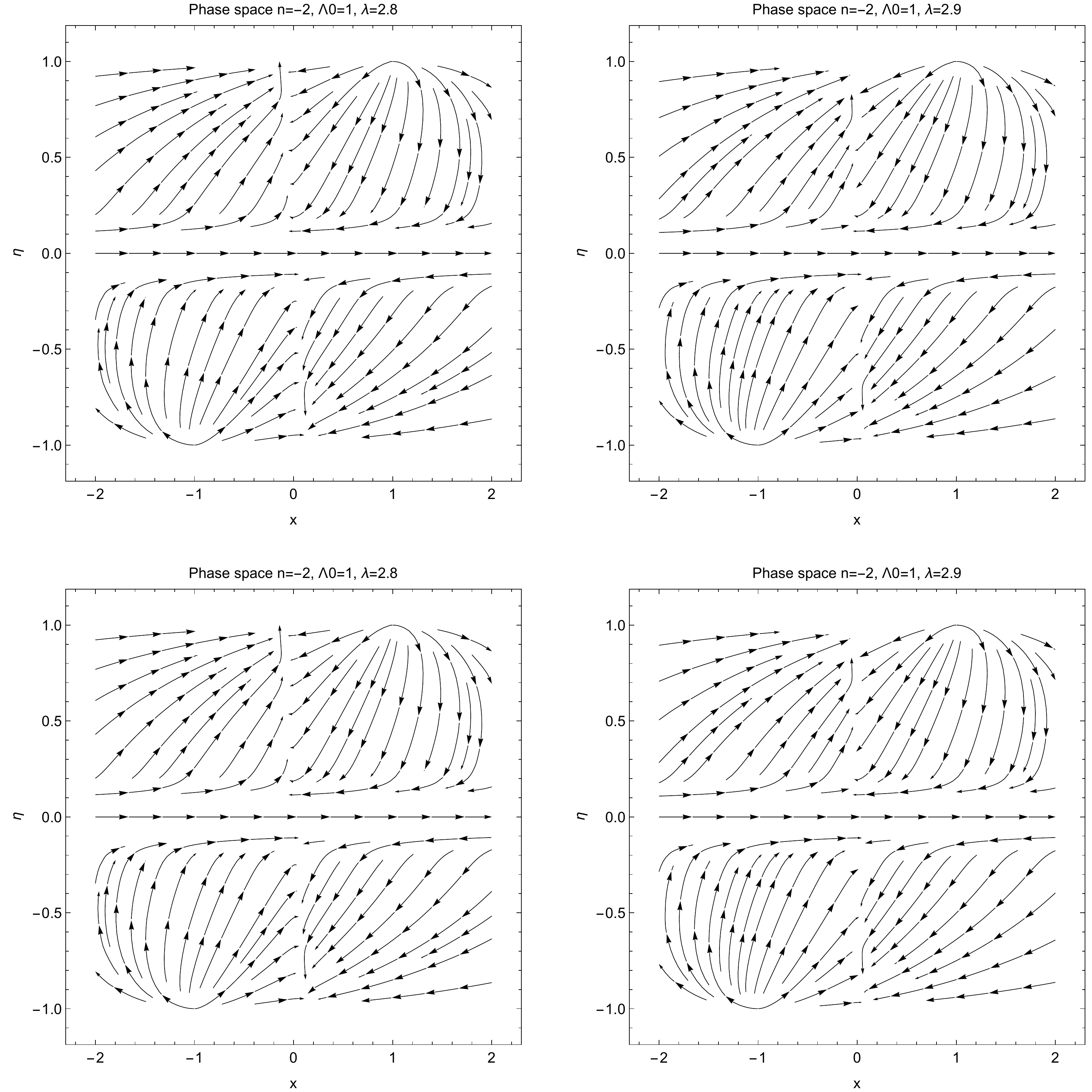} \caption{Phase-space
portrait for the two-dimensional dynamical system in the variables $\left\{
x,\eta\right\}  $ for the values of the free parameters described in Fig.
\ref{ftb2}.}%
\label{ftb3}%
\end{figure}

Specifically, in Figs. \ref{ftb2} we present the qualitative evolution of the
physical parameter $w_{eff}$ and of $x\left(  \omega\right)  ,~\eta\left(
\omega\right)  $ for different values of the variable $\lambda$ and for
$n=-2$,~$\Lambda_{0}=1$.\ The phase space portraits are given in Fig.
\ref{ftb3}. It is clear that for $\lambda<3,$ the final attractor is point
$P_{3},~$while for $\lambda>3$ the final attractor is the de Sitter points
$P_{5}$. For $n<0$, it is clear from the phase space that the line $\eta=0$
can not be crossed.

\begin{figure}[ptb]
\centering\includegraphics[width=1\textwidth]{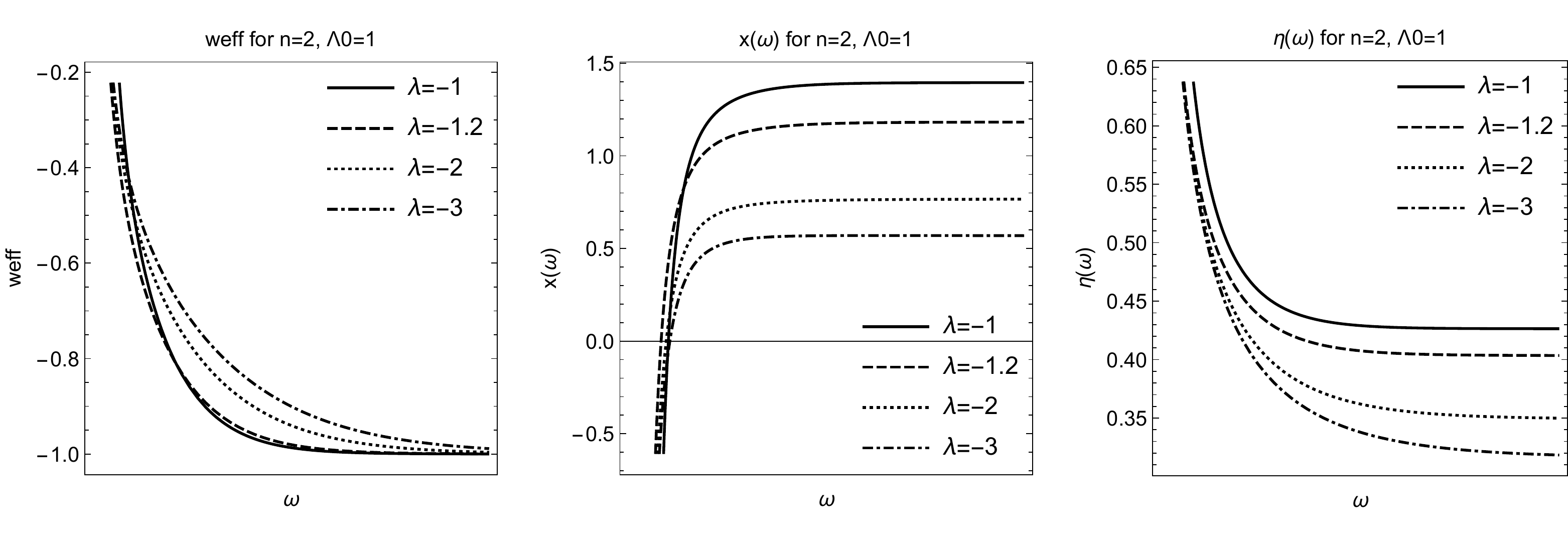} \caption{Qualitative
evolution for the physical variable $w_{eff}$ and for the parameters $x,~\eta$
for $n=2,~\Lambda_{0}=1$ and for different values of $\lambda$. We observe
that the final attractor is always the de Sitter universe of point $P_{5}$}%
\label{ftb4}%
\end{figure}

Similarly, for $n=-2$, and for $\Lambda_{0}=1$, the qualitative evolution of
the physical parameters and the phase-space portraits are presented in Figs.
\ref{ftb3} and \ref{ftb4}. There it is clear that the final attractor is
always the de Sitter universes described by $P_{5}$. From the plots, it is
clear that the Minkowski point $P_{6}$ is a saddle point.

From the phase-space diagrams it is clear that there are trajectories which take values at the infinity for the variable $x$. In the following, we present the the analysis for the existence of stationary points at the infinity.

\begin{figure}[ptb]
\centering\includegraphics[width=1\textwidth]{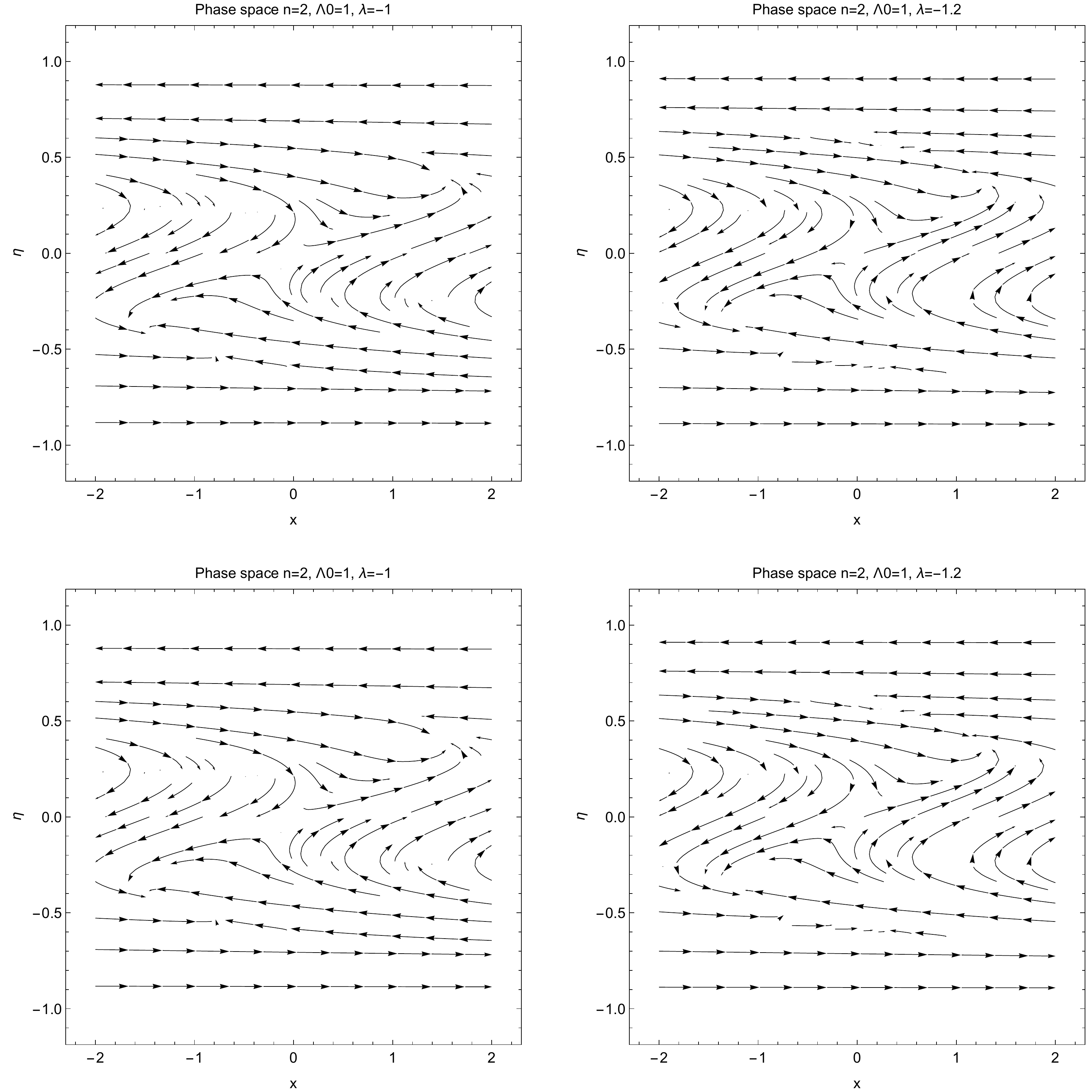} \caption{Phase-space
portrait for the two-dimensional dynamical system in the variables $\left\{
x,\eta\right\}  $ for the values of the free parameters described in Fig.
\ref{ftb4}.}%
\label{ftb5}%
\end{figure}

\subsection{Analysis at infinity}

We continue by studying the evolution of the system at infinity. Parameter
$\eta$ is bounded, but that is not true for parameter $x$ which can take
values at infinity. This part of the investigation has not been performed
before in \cite{cel1}.

We do the change of variables $x=\frac{X}{\sqrt{1-X^{2}}}$, $d\varpi
=\sqrt{1-X^{2}}d\omega$, where we study the dynamical system for $X=\pm1$. In
the following we consider the case $X=1$. In the new variables the field
equations are written as%
\begin{align}
\frac{dX}{d\varpi}  &  =0~,~\label{ft.17}\\
\frac{d\eta}{d\varpi}  &  =-\lambda\eta^{2}\left(  1-\eta^{2}\right)  .
\label{ft.18}%
\end{align}

Hence, the stationary points are for $\eta=\pm1$ and $\eta=0$. For $\eta=\pm
1$, from the second equation we derive that near the stationary point
$\frac{d\eta}{d\varpi}|_{\eta\rightarrow1}=\pm2\lambda$. On the other hand,
for the Minkowski points we derive $\frac{d\eta}{d\varpi}=0$. Furthermore, can
be solved by quadratures. \ We perform the second change of variable
$d\varpi=\eta d\sigma$, that is, from (\ref{ft.18}) it follows
\begin{equation}
\eta\left(  \sigma\right)  =\left(  1+e^{2\lambda\sigma}\right)  ^{-1}\text{.
} \label{ft.19}%
\end{equation}
Hence, for $\lambda>0$, $\eta=0$ is a future attractor while for $\eta<1$,
$\eta=1$ is a future attractor, for equation (\ref{ft.18}).

\begin{figure}[ptb]
\centering\includegraphics[width=1\textwidth]{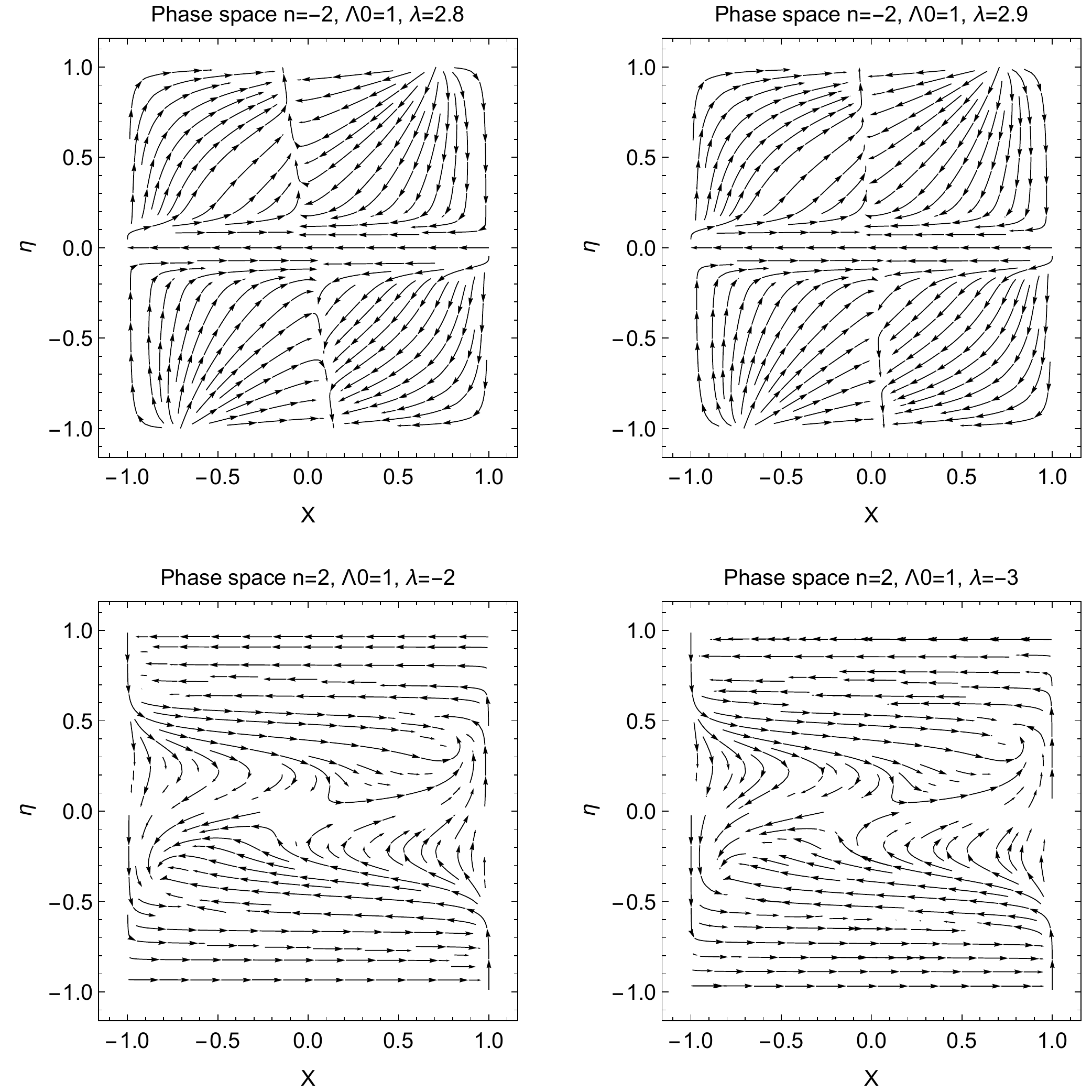} \caption{Phase-space
portrait for the two-dimensional dynamical system (\ref{ft.17}), (\ref{ft.18})
in the variables $\left\{  X,\eta\right\}  $. We observe that at the
infinity,~$X^{2}=1$, the stationary points are saddle points or sources.
Attractors exist only in the finite regime, $X^{2}<1$. The phase-space
portraits are for values of the free parameters as presented in Figs.
\ref{ftb3} and \ref{ftb5}.}%
\label{ftb6}%
\end{figure}

\section{Conclusions}

\label{sec4}

The determination of the stationary points and the study of their stability is
an important approach in order to understand the cosmological evolution and
the validity of a specific theory of gravity. In this work we investigated the
global dynamics for the fourth-order teleparallel $f\left(  T,B\right)  $
theory of gravity in a spatially flat FLRW background space. The approach that
we applied is more general than that of the Hubble normalization, where we
found that Minkowski spacetime can be an exact solution for the field equation
described by a stationary point.

In this analysis firstly we wrote the second-order field equations into a
system of algebraic-differential equation with the use of new dimensionless
variables. The stationary points of the algebraic-differential system were
investigated in the local variables as also in the infinite regime. For the
functional for of $f\left(  T,B\right)  $ we assumed $f\left(  T,B\right)
=F\left(  T\right)  +\beta\left(  B\right)  $, in which $F\left(  T\right)
=T+\Lambda_{0}T^{n}$ and $\beta\left(  B\right)  =-\frac{B}{\lambda}\ln\left(
B\right)  $. These two functions reduce the algebraic-differential system into
a system of two first-order ordinary differential equations. From previous
studies it is known that such $\beta\left(  B\right)  $ function provides
dynamical terms in the field equations that can describe important areas of
the cosmological history~\cite{angrg}. Furthermore, for $B=B_{0}>0$,
$\beta\left(  B\right)  $ is approximated as $\beta\left(  B\right)
\simeq\left(  B-B_{0}\right)  +\frac{1}{2}\left(  B-B_{0}\right)  ^{2}%
-\frac{1}{3!}\left(  B-B_{0}\right)  ^{3}+...$. Furthermore, function
$F\left(  T\right)  =T+\Lambda_{0}T^{n}$ has been widely studied in the
literature before as a dark energy candidate and there are many applications
in the literature \cite{Ferraro}.

In the finite regime, that is, in the local variables we find that the
cosmological field equations admit six stationary points. Points $P_{1}%
,~P_{2}$,~describe asymptotic solutions where the effective fluid is that of
the stiff fluid. Points $P_{3}$ and $P_{4}$ correspond to ideal gas solutions,
or to the de Sitter universe for $\lambda=3$. Point $P_{5}$ describes a new
family of de Sitter universes, while $P_{6}$ corresponds to the Minkowski
universe which is always a saddle point. For these points only one of the
points $P_{3}$,$~P_{4}$ or $P_{5}$ can be an attractor. That means, that the
model can describe a cosmological evolution with at least two expansion eras,
a scaling solution and an de Sitter universe, or we can say that it can
describes the late time acceleration phase of the universe and an additional
matter epoch. It is clear that in order to have all the complete cosmological
history, such as radiation and matter eras additional matter source should be
considered. However, from these results it is clear that this cosmological
model is viable. In addition, we studied the dynamics at the infinite, where
we found that no new attractors exists.

This analysis supports the statement that in terms of the background equations
$f\left(  T,B\right)  $ can play an important role for the description for
some important eras of the cosmological history. In a future study we plan to
investigate the effects of the curvature term in the background space to the
field equations.

\begin{acknowledgments}
The research of AP and GL was funded by Agencia Nacional de Investigaci\'{o}n
y Desarrollo - ANID through the program FONDECYT Iniciaci\'{o}n grant no.
11180126. Additionally, GL was funded by Vicerrector\'{\i}a de
Investigaci\'{o}n y Desarrollo Tecnol\'{o}gico at Universidad Cat\'{o}lica del Norte.
\end{acknowledgments}

\end{document}